# Observation of Coexisting Dirac Bands and Moiré Flat Bands in Magic-Angle Twisted Trilayer Graphene

Yiwei Li, Shihao Zhang, Fanqiang Chen, Liyang Wei, Zonglin Zhang, Hanbo Xiao, Han Gao, Moyu Chen, Shijun Liang, Ding Pei, Lixuan Xu, Kenji Watanabe, Takashi Taniguchi, Lexian Yang, Feng Miao, Jianpeng Liu*, Bin Cheng*, Meixiao Wang*, Yulin Chen*, and Zhongkai Liu*

Y. Li, S. Zhang, L. Wei, Z. Zhang, H. Xiao, H. Gao, D. Pei, J. Liu, M. Wang, Y. Chen, Z. Liu

School of Physical Science and Technology

ShanghaiTech University

Shanghai 201210, P. R. China

Email: liuzhk@shanghaitech.edu.cn; wangmx@shanghaitech.edu.cn; liujp@shanghaitech.edu.cn

Y. Li

Institute for Advanced Studies

Wuhan University

Luojiashan Street, Wuhan, Hubei, 430072, P. R. China

S. Zhang, D. Pei, J. Liu, M. Wang, Y. Chen, Z. Liu

ShanghaiTech Laboratory for Topological Physics

Shanghai 200031, P. R. China

F. Chen, M. Chen, S. Liang, F. Miao

National Laboratory of Solid State Microstructures, School of Physics

Institute of Brain-Inspired Intelligence, Collaborative Innovation Center of Advanced Microstructures

Nanjing University

Nanjing 210093, P. R. China

L. Xu, L. Yang, Y. Chen

State Key Laboratory of Low Dimensional Quantum Physics

Department of Physics

Tsinghua University

Beijing 100084, P. R. China

K. Watanabe

Research Center for Functional Materials

National Institute for Materials Science, 1-1 Namiki

Tsukuba 305-0044, Japan

T. Taniguchi

International Center for Materials Nanoarchitectonics

National Institute for Materials Science, 1-1 Namiki

Tsukuba 305-0044, Japan

L. Yang

Frontier Science Center for Quantum Information

Beijing 100084, P. R. China

B. Chen

Institute of Interdisciplinary Physical Sciences

School of Science, Nanjing University of Science and Technology

Nanjing 210094, P. R. China

Email: bincheng@njust.edu.cn

Y. Chen

Clarendon Laboratory

Department of Physics

University of Oxford

Oxford OX1 3PU, U.K

Email: yulin.chen@physics.ox.ac.uk

**Moiré superlattices that consist of two or more layers of two-dimensional materials stacked together with a small twist angle have emerged as a tunable platform to realize various correlated and topological phases, such as Mott insulators, unconventional superconductivity and quantum anomalous Hall effect. Recently, the magic-angle twisted trilayer graphene (MATTG) has shown both robust superconductivity similar to magic-angle twisted bilayer graphene (MATBG) and other unique properties, including the Pauli-limit violating and re-entrant superconductivity. These rich properties are deeply rooted in its electronic structure under the influence of distinct moiré potential and mirror symmetry. Here, combining nanometer-scale spatially resolved angle-resolved photoemission spectroscopy (nano-ARPES) and scanning tunneling microscopy/spectroscopy (STM/STS), we systematically measure the yet unexplored band structure of MATTG near charge neutrality. Our measurements reveal the coexistence of the distinct dispersive Dirac band with the emergent moiré flat band, showing nice agreement with the theoretical calculations. These results serve as a stepstone for further understanding of the unconventional superconductivity in MATTG.**

# 1. Introduction

Artificial moiré superlattices via stacking van der Waals materials with a twist angle and/or a lattice mismatch have emerged as a unique platform with unprecedented control to investigate new correlated and topological phases beyond their building blocks.[1-22] As an archetypical example, magic-angle (~1.1°) twisted bilayer graphene (MATBG) hosts rich quantum phases, including correlated insulating state,[1] unconventional superconductivity,[2] strange metal behavior,[21] nematicity,[22] ferromagnetism,[14] and quantum anomalous Hall effect[15] under distinct band-filling and electrical field conditions. These intriguing properties originate from the flat bands near charge neutrality[1,2,23,24] whereas the Dirac band in the monolayer graphene is highly dispersive with a typical Fermi velocity of $10^6$ *m/s*. The original Dirac bands from the two graphene monolayers are folded into the mini-Brillouin zone (mini-BZ) by the long-wavelength moiré superlattice potential in MATBG with interlayer hybridization,[25,26] leading to the distinct flat band structure.

With three graphene layers with alternating twist angles $\theta$ and $-\theta$ (~1.6°) stacked together, magic-angle twisted trilayer graphene (MATTG) has recently emerged as a versatile platform for studying correlated physics and unconventional superconductivity.[17-19] MATTG shares mathematically equivalent flat bands with MATBG with renormalized magic angle,[27] which give rise to a similar phase diagram with a superconducting dome between correlated regions around integer flat-band fillings.[1,2,17,19] However, compared with MATBG, MATTG exhibits unique properties such as "ultrastrong" coupling superconductivity in proximity to BEC-BCS crossover,[17] Pauli-limit violated critical in-plane magnetic field and re-entrant superconductivity behavior.[18] These unique properties can be traced to its distinct electronic band structure with coexisting flat bands and highly dispersive Dirac bands that can be hybridized with each other under mirror-symmetry-breaking electric displacement field.[28]

The extraordinary phenomena of the moiré graphene superlattices could be attributed to the emergent moiré flat bands in the mini-BZ. Therefore, the characterization of the moiré bands structure is a critical yet challenging mission from both theoretical and experimental perspectives. From the theoretical perspective, it is demanding to calculate the electronic structure fully from first principles for a moiré unit cell containing thousands of atoms, thus effective low-energy models and various approximations need to be adopted. Experimentally, common approaches to probe the electronic band structures include the Landau fan diagram,[17,29] which requires sample homogeneity and a priori knowledge of the band structure; and the scanning tunneling spectroscopy (STS),[30] which lacks momentum-space resolution. Therefore, direct observation of the electronic band dispersions in reciprocal space could provide essential information to the understanding of the unique moiré physics in moiré graphene systems.

In this work, we directly visualize the electronic band dispersion in the MATTG device by nanometer-scale spatially resolved angle-resolved photoemission spectroscopy (nano-ARPES), which effectively avoids the possible surface inhomogeneity. The fine momentum resolution (~ 0.01 Å$^{-1}$) that is significantly smaller than the moiré reciprocal vector (~ 0.08 Å$^{-1}$), enables us to resolve the details of the moiré mini-bands. Our results confirm the coexistence of the dispersive Dirac bands and the moiré flat bands as predicted by theoretical calculation[27,31,32] and consistent with previous transport and STM/STS studies;[30] the discovered moiré flat bands extend over an entire mini-BZ, reflecting their localization in the moiré lattice. Our results provide a firm basis for further investigation of the rich strongly correlated physics and unconventional superconductivity in MATTG.

## 2. Experimental Results

Figure 1a,b shows the side view of the device configuration and measurement geometry, respectively. The MATTG device is fabricated via the tear-and-stack method[17] (see Experiemtal Section) with a controlled

twist angle near 1.6°. To achieve decent momentum resolution, a hexagonal boron nitride (hBN) flake on the Si/SiO$_2$ substrate provides an atomically flat surface that supports the MATTG sample. The sample is grounded via graphite and gold contact for both ARPES and STM/STS measurements.

Figure 1c shows the optical image of the device and Figure 1d shows the corresponding spatial map of integrated nano-ARPES intensity near the Fermi level, which allows us to identify different regions in the device, including the MATTG, twisted bilayer graphene (TBG) and monolayer graphene (MG) region. The STM measurement characterizes the MATTG region which shows a moiré superlattice wavelength of 8.75 nm (see Figure 1e), corresponding to the magic twist angle of 1.6°. Characteristic nano-ARPES spectra collected at MATTG, TBG and MG regions are presented in Figure 1f, respectively. Signatures of intricate moiré-potential-induced mini-bands are observed in MATTG (see Figure 1f(i)) and TBG (see Figure 1f(ii)). A slightly p-doped Dirac cone dispersion of the MG region (see Figure 1f(iii)) is due to the charge transfer from the hBN layer, consistent with previous study.[33]

The STM topography shown in Figure 1e suggests that the alternating trilayer graphene system has the energetically favorable A-tw-A stacking configuration ('tw' denotes the middle twisted layer) without top-bottom lattice translational shift,[30,31,34] as illustrated in Figure 2a, Such a configuration possesses mirror symmetry that is absent in MATBG. In real space, the two reversed twists from three graphene layers results in alternating AAA, ABA and ACA regions within the moiré superlattice (see Figure 2(a)).[17,30,31,34] In the momentum space, without the interlayer coupling and the resulted long-wavelength moiré potential, the low-energy band structure of MATTG can be regarded as three sets of monolayer graphene Dirac cones at K$_{1,3}$ and K$_2$ of two BZs rotated about $\Gamma$ by $\theta$ (the top and bottom graphene layers share the same BZ, see Figure 2b). The moiré periodicity in the real space corresponds to the mini-BZ as shown by the orange hexagon in Figure 2b with a reciprocal vector of $|\boldsymbol{G}_{moiré}| = 4\pi/\sqrt{3}L_{moiré}$. With interlayer coupling, the moiré superlattice geometry would fold the band structures from the original BZ into the mini-BZ by integer multiples of $G_{moiré}$, while the moiré potential leads to the hybridizations between the three Dirac cones.[1,25,26] Although the number of bands increases dramatically in the mini-BZ as the unit cell has been expanded by over a thousand times in real space, the spectral weights of the low-energy bands are still mainly concentrated near the $K_{1,3}$ and $K_2$ points of the original BZs, as shown in Figure 2c,d. The spectral intensity of the moiré mini-bands can be interpreted as a measure of the long-wavelength modulation of charge density induced by the moiré potential, which lowers the translational symmetry of the original crystal periodicity, similar to the scenarios of charge-density-wave materials.[35-38]

Simulated constant energy surfaces at $E_F$ - 0.8 eV and $E_F$ - 1 eV show intricate textures consisting of multiple rounded triangular pockets nested with each other (see Figure 2c). The corresponding nano-ARPES results shown in Figure 2d are generally consistent with the calculations (also see Figure S2, Supporting Information), but are strikingly different from those of graphene multilayers with large twist angles, in which the energy contours are weakly hybridized rounded triangular pockets centered at the corresponding K points.[39] One might notice that only a part of features with similar spectral weight in the calculations have been experimentally resolved, which we attribute to the matrix element effect and a finite photoelectron mean-free-path.[40] The latter factor results in strong spectral weight from the top layer and much weakened signals from the layers below. The pocket centered at $K_{1,3}$, as indicated by the blue arrow in Figure 2c,d, contributes the most significant spectral weight, which originates from the Dirac cone of the top and bottom graphene layers.[41] This dispersive Dirac band survives the moiré potential under the protection of the mirror symmetry,[27,31,32] as we will elaborate in Figure 3. Interestingly, we notice that the momentum separation between neighboring pockets matches well with the moiré reciprocal vector $G_{moiré}$, as particularly shown in the zoom-in momentum region (see Figure 2e). The momentum-distribution-curve (MDC) plot (see Figure 2f) indicates a peak-to-peak distance of 0.083±0.01 Å, which matches $G_{\text{moiré}}$ with a twist angle of 1.6°±0.2° near the theoretical magic angle. The nice agreement between the observed pattern and the calculations provides direct experimental evidence for the moiré potential effects on the electronic structure of MATTG in momentum space.

Now we focus on visualizing the characteristic low-energy electronic structures in MATTG, including both the dispersive Dirac bands and flat bands, as illustrated in Figure 3. The low-energy electronic structure of MATTG shares some similarities with that of MATBG, hosting flat bands across the entire mini-BZ and a band top at the mini-BZ center γ. Compared with the MATBG, the distinctive electronic structure of the MATTG is the gapless Dirac cone at $K_{1,3}$ (see Figure 3a). Spectral weight simulation (see Figure 3b) shows consistent band signatures as indicated by the continuum model calculation (see Figure 3a), although the band gap (between the flat bands and the remote energy bands) at γ might be beyond the energy resolution of 35 meV (see Experimental Section). Constant energy ARPES mapping integrated from $E_F$ – 50 meV to $E_F$ (see Figure 3d) demonstrates discernible spectral weight from the Dirac point at $K_{1,3}$ and the band top at γ, consistent with the corresponding simulation based on the atomistic tight-binding model in Figure 3c (see Experimental Section). The highly dispersive Dirac cone and the flat bands are also directly visualized via various momentum cuts across $K_{1,3}$ shown in Figure 3e-g. The Fermi velocity of the Dirac band is estimated to be $(1\pm0.04) \times 10^6$ *m/s*, comparable to the typical value of monolayer graphene, indicating negligible interaction with the flat band. The intricate texture of moiré mini-bands is also presented in these momentum cuts, showing qualitative consistency with the calculation based on the atomistic tight-binding model (see Experimental Section).

The flat band can be clearly seen in the nano-ARPES cuts (see Figure 4b,e) along two perpendicular momentum directions integrated in a 0.3×0.3 Å$^{-2}$ region near $K_{1,3}$ and $K_2$, as indicated by the grey shaded square in Figure 4 a,d. Energy-distribution-curve (EDC) plots (see Figure 4c,f) reveal non-dispersive peak features near $E_F$ extending over a large momentum area (FWHM ≈ 0.2 Å$^{-1}$, see Figure S3, Supporting Information) that exceeds moiré wavelength ($G_{moiré}$ ≈ 0.08 Å$^{-1}$). The extension of the flat band in the momentum space indicates the localization of the flat-band wavefunctions in real space.[2,42] The STS intensity near $E_F$ is mostly concentrated in the AAA region (FWHM ≈ 5.3 nm, see Figure 4h and Figure S3, Supporting Information), whereas small but non-zero amplitudes on the ABA/ACA regions give rise to the weak dispersion of the flat band, similar to MATBG.[2,42-44] ABA/ACA regions also show prominent V-shape spectra away from $E_F$ (see Figure 4i), which originates from the Dirac cone, consistent with previous STM/STS study on MATTG.[30] The double-peak feature with an energy splitting of 25-55 meV (see Figure 4i and Figure S4, Supporting Information) might result from the $C_3$-symmetry-break strain, lattice relaxation effects and/or electron correlations, as commonly observed in MATBG[43-48] and MATTG.[34] The consistent low-energy spectra collected by nano-ARPES and STS, as summarized in Figure 4i, further demonstrate the coexistence of flat bands and dispersive Dirac bands.

## 3. Discussions and Outlook

STM/STS measurements focus on the electronic structure with atomic precision and high energy resolution, whereas nano-ARPES measurements with momentum resolution ability are comparably more global, which covers a spatial range of about 400 nm. By performing STM topography measurements on different positions of the sample, we estimated that the twist angle variance of the sample is about 0.06° in a spatial range of several microns (see Figure S6, Supporting Information). Since the twist angle inhomogeneity is significantly smaller than the corresponding nano-ARPES momentum resolution ability (±0.2°), it is convincing to make direct comparisons among the nano-ARPES spectra, integrated STS and theoretical calculations.

In addition to the well-studied MATBG, our combined ARPES and STM study provides the groundwork of direct visualization of the coexisting dispersive Dirac bands, flat bands and moiré mini-bands in MATTG. Under the experiment resolution, no hybridization gap is observed between the Dirac band and flat bands (see Figure 3), in contrast to the hybridization gap observed in MATBG.[24] Applying a vertical displacement field would break the mirror symmetry of the system, which would induce hybridizations between the flat bands and the Dirac bands.[17] The additional mirror symmetry and the mutual interactions between the strongly correlated electrons in the flat bands and the itinerant Dirac fermions in MATTG may be crucial in understanding the unconventional superconductivity and other correlated phases that are distinct from those observed in MATBG. First, the Dirac band in the MATTG has a large bandwidth, whose kinetic energy prevails over the electron-electron Coulomb interaction and protects the Dirac cone from interaction instabilities, making MATTG metallic even at full/empty filling of the flat bands.[49] Second, in MATTG, the combined mirror and $C_{2z}$ symmetry preserves under finite in-plane magnetic fields[28] and retains the degeneracy between states of opposite valleys/moiré wavevectors, supporting a valley-singlet, spin-triplet superconductivity even in the presence of in-plane magnetic fields.[28] Lastly, the coexisting localized flat bands and the highly dispersive Dirac bands in MATTG are reminiscent of the heavy-fermion problem in the rare-earth element compounds, in which the mutual interactions between the local magnetic moments and the itinerant electron degrees of freedom give rise to a variety of unusual phenomena such as quantum criticality and unconventional superconductivity.[50-52]

## 4. Summary

In conclusion, we have presented a systematic investigation of the low energy electronic structure of the moiré superlattice system MATTG. We observed moiré induced mini-bands with a momentum separation consistent with the magic twist angle. We further observed the coexistence of moiré flat band and dispersive Dirac band, the latter being guaranteed by the mirror symmetry which is absent in MATBG. The extension of the flat band over an entire mini-BZ in the momentum space, as manifested by our nano-ARPES results, is consistent with its localization near the AAA site in the real space, as indicated by our STM measurements. our observations establish the pivotal features in the band structure for further exploration of the unconventional superconductivity and other correlated states in MATTG.

## 5. Experimental Section

*Sample Fabrication:* To obtain the uncapped MATTG, we adopted a dry transfer method followed by flipping the heterostructure. We exfoliated graphene and hBN flakes on $SiO_2$/Si substrates, and the three graphene layers were cut from single graphene flakes by scanning probe nanolithography. In the first stage, an hBN-MATTG structure was assembled via the dry transfer method, where a PC/PDMS stamp was used to pick the hBN at 80 °C and the graphene flakes at room temperature sequentially. Then the stack was released on the $SiO_2$/Si substrate at 140 °C. After dissolving the PC film in chloroform, the stack was spin-coated with a thin polypropylene carbonate (PPC) film. Subsequently, the PPC film, as well as the hBN-MATTG structure, was peeled off from the substrate, then flipped over and transferred to a new $SiO_2$/Si substrate. Finally, we removed the polymer through annealing in an argon ambient at 300 °C for 3 hours. The electrical contacts were thermally evaporated with Cr 5 nm / Au 40 nm after patterned by electron beam lithography (EBL).

*Nano-ARPES Measurements:* The nano-ARPES measurements were performed at the BL07U endstation of Shanghai Synchrotron Radiation Facility (SSRF). The base pressure is lower than $5 \times 10^{-11}$ mbar. The sample was annealed at 200 °C for 12 hours prior to the measurement to desorb absorbates. The beam was focused using a Fresnel zone plate (FZP) and a spatial resolution of ~ 400 nm was achieved. All the experiments were performed at $T$ = 40 K with a photon energy of 92 eV and linear-horizontal polarized light with a fixed incidence angle of 60°. The data were collected by a hemispherical Scienta DA30 electron analyzer. The instrumental energy and momentum resolution under the measurement conditions was ~35 meV and 0.01 Å, respectively.

*STM/STS Measurements:* The sample was annealed at 200 °C for 12 hours in the ultrahigh vacuum preparation chamber before the STM/STS experiment. After it was cooled down to room temperature, it was then transferred to a cryogenic stage to keep it at 5 K. Pt-Ir tips had already been calibrated on the silver islands grown on p-type Si (111) 7×7 before they were used for imaging and tunneling. A combination of lenses and CCD was used to localize the sample area and help the STM tip approach it. The Lock-in technique with a 5 mV modulation at 971.33 Hz was used to obtain the *dI/dV* curve.

*Continuum Model Calculations:* We use the continuum model to describe the low-energy electronic structure of twisted trilayer graphene based on the Bistritzer-MacDonald model.[25] In $\mu$ valley, the Hamiltonian of continuum model can be expressed as

$$H_\mu(r) = \begin{bmatrix} -\hbar v_F(k - K_1^\mu) \cdot \sigma^\mu & U_\mu^+(r) & 0 \\ U_\mu(r) & -\hbar v_F(k - K_2^\mu) \cdot \sigma^\mu & U_\mu(r) \\ 0 & U_\mu^+(r) & -\hbar v_F(k - K_3^\mu) \cdot \sigma^\mu \end{bmatrix}$$

where $v_F$ refers to Fermi velocity and $\sigma^\mu = (\mu\sigma_x, \sigma_y)$ is the Pauli matrix in the sublattice space. Our model is expanded near the Dirac point of three layers $K_l^\mu (l = 1,2,3)$, and we use the $U_\mu(r)$ to describe the interlayer coupling between layers.

$$U_\mu(r) = \begin{bmatrix} u_0 g_\mu(r) & u_0' g_\mu(r - \mu r_{AB}) \\ u_0' g_\mu(r + \mu r_{AB}) & u_0 g_\mu(r) \end{bmatrix} e^{i\mu\Delta K \cdot r}$$

in which $r_{AB} = (L_s/\sqrt{3}, 0)$ and $L_s$ is the length of the moiré lattice vector. $\Delta K = (0, 4\pi/3L_s)$ is the shift between the Dirac points of layers due to the twist. In this moiré potential, we adopt the phase factor $g_\mu(r) = \sum_{j=1}^{3} e^{-i\mu q_j \cdot r}$ with $q_1 = (0, 4\pi/3L_s)$, $q_2 = (-2\pi/\sqrt{3}L_s, -2\pi/3L_s)$ and $q_3 = (2\pi/\sqrt{3}L_s, -2\pi/3L_s)$. Because of the atomic corrugations, intrasublattice interlayer tunneling $u_0 = 0.0797$ eV is smaller than

intersublattice interlayer tunneling $u_0^{'} = 0.0975 eV$.[41]

*Atomistic Tight-Binding Model Calculations:* As we need to study the energy bands far away from charge neutrality, we also use a realistic atomic tight-binding model to study the electronic structure of MATTG. In the atomistic tight-binding model, we use an empirical Slater-Koster type parameter $t(\boldsymbol{d})$ to describe the hopping between the $p_z$ orbitals at different carbon sites.[53]

$$t(\boldsymbol{d}) = V_\sigma^0 e^{-(r-d_c)/\delta_0} \left(\frac{\boldsymbol{d}\cdot\hat{\boldsymbol{z}}}{d}\right)^2 + V_\pi^0 e^{-(r-a_0)/\delta_0}\left[1-\left(\frac{\boldsymbol{d}\cdot\hat{\boldsymbol{z}}}{d}\right)^2\right]$$

where $V_\sigma^0 = 0.48$ eV, $V_\pi^0 = -2.7$ eV, and $d_c$ is 3.35 Å representing the interlayer distance Bernal bilayer graphene. $\delta_0 = 0.184\, a$, where $a$ is the lattice constant of graphene and $a_0 = a/\sqrt{3}$. $\boldsymbol{d}$ is the displacement vector between two carbon sites. There are typically atomic corrugations in the twisted graphene systems, in which the interlayer distance varies at different regions in the moiré superlattice. Here we assume the interlayer distance varies as a cosine function of in-plane real space **r** with moiré periodicity,[41]

$$d_z = d_0 + 2d_1 \sum_{j=1}^{3} \cos(\boldsymbol{g}_i \cdot \boldsymbol{r}),$$

where $\boldsymbol{g}_1$, $\boldsymbol{g}_2$ and $\boldsymbol{g}_3 = \boldsymbol{g}_1 + \boldsymbol{g}_2$ are the three primitive reciprocal lattice vectors of the moiré supercell. And we set $d_0 = 3.433$ Å and $d_1 = 0.0278$ Å in order to reproduce the interlayer distances of the AA-stacked graphene and AB-stacked graphene in the AA site and AB/BA site, respectively.[41]

*Spectral Function Calculations:* To simulate the ARPES intensity, we calculate the unfolded spectral functions for the MATTG system along **k** path in the atomic BZ near the K valley,[54] and the unfolded spectral function is expressed as

$$A(\omega, \boldsymbol{k}) = -\frac{1}{\pi}\text{Im}\left[\sum_{l\alpha}\sum_n \frac{|\langle l\alpha\boldsymbol{k}|n\boldsymbol{k}\rangle|^2}{\omega - E_{n\boldsymbol{k}} + i\delta}\right]$$

where $l$ refers to the layer index and $\alpha$ represents the sublattice index. $|n\boldsymbol{k}\rangle$ is the Bloch wave function at the $\boldsymbol{k}$ point of the $n$-th moiré energy band, and $|l\alpha\boldsymbol{k}\rangle$ denotes the Bloch wave function of the $l$th monolayer graphene and $\alpha$ sublattice, without the moiré potential effects. The $n$th moiré Bloch function at wavevector $k$ can be expressed as:

$$|n\mathbf{k}\rangle = \sum_{il\alpha} C_{il\alpha,n}(\mathbf{k})|il\alpha\mathbf{k}\rangle$$

where $i$ stands for the atomic lattice vector index within a moiré primitive cell., The Bloch-like state $|il\alpha\mathbf{k}\rangle$ at wavevector $k$ can be expressed on the basis of the $p_z$-orbital-like Wannier functions as:

$$|il\alpha\mathbf{k}\rangle = \frac{1}{\sqrt{N_M}}\sum_{\mathbf{R}} e^{i\mathbf{k}\cdot\mathbf{R}}|il\alpha\mathbf{R}\rangle,$$

where $\mathbf{R}$ is the moiré lattice vector. Then we obtain

$$\langle l\alpha\mathbf{k}|n\mathbf{k}\rangle = \sum_{\mathbf{R}i'l'\alpha'} \langle l\alpha\mathbf{k}|i'l'\alpha'\mathbf{R}\rangle\langle i'l'\alpha'\mathbf{R}|n\mathbf{k}\rangle = \frac{1}{\sqrt{N_0}}\sum_i e^{-i\mathbf{k}\cdot\mathbf{a}_i} C_{il\alpha,n}(\mathbf{k})$$

in which $\mathbf{a}_i$ refers to the $i$th atomic lattice vector within a moiré primitive cell for the $l$-th layer, and $N_0$ is the number of atomic unit cells within a moiré primitive cell.

## Supporting Information

Supporting Information is available from the Wiley Online Library or from the author.

## Acknowledgements

Y.L., S.Z., F.C and L.W. contributed equally to this work. Z.L. acknowledges the National Key R&D program of China (Grant No. 2017YFA0305400). B.C. and F.M. acknowledge the National Natural Science Foundation of China (12074176, 62122036, 62034004, 61921005, 61974176), the Strategic Priority Research Program of the Chinese Academy of Sciences (XDB44000000), and F.M. acknowledges the support from the AIQ foundation. Y.L. acknowledges the support from the International Postdoctoral Exchange Fellowship Program (Talent-Introduction Program, Grants No. YJ20200126), the fellowship of China Postdoctoral Science Foundation (2021M692131) and the National Natural Science Foundation of China (12104304). S.Z. and J.L. acknowledge National Key R&D program of China (Grant No. 2020YFA0309601), the National Natural Science Foundation of China (Grant No. 12174257), and the start-up grant of ShanghaiTech University.

## Conflict of Interest

The authors declare no conflict of interest.

## Data Availability Statement

The data that support the findings of this study are available from the corresponding author upon reasonable request.

## Keywords

Magic-angle twisted trilayer graphene, Dirac fermion, flat bands, moiré pattern, nano-ARPES

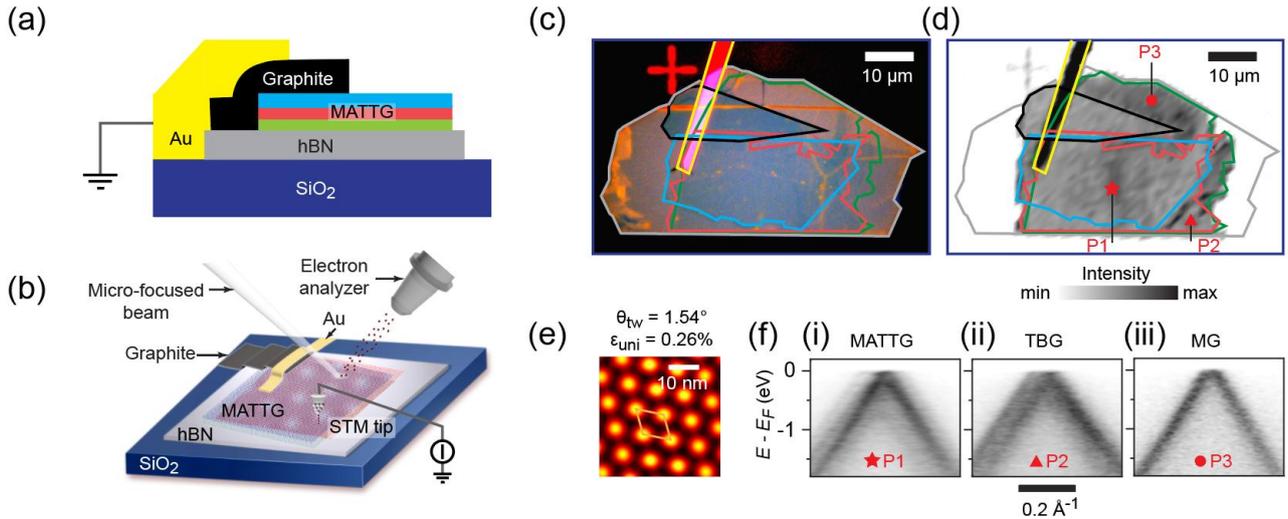

**Figure 1**. MATTG device configuration. a,b) Side-view schematics of the MATTG/hBN/SiO$_2$ sample (a) and 3D illustration of the measurement geometry (b). Each area is indicated as follows: dark blue, SiO$_2$; yellow: gold contact; black: graphite electrode; grey: hBN; green: bottom graphene monolayer; red: middle graphene monolayer; blue: top graphene monolayer. c,d) Optical micrograph (c) and scanning photoemission microscopy (SPEM) image of the sample (d). The boundary of each segment is indicated with the same color as in (s) and (b). e) STM Topography of the MATTG area. Set voltage $V_{set}$ = 100 mV and set current $I_{set}$ = 250 pA. The twist angle and strain are indicated according to the moiré wavelengths (see Figure S6, Supporting Information). f) The energy-momentum band dispersion around the K point of MATTG (i), TBG (ii) and MG (iii) areas. These spectra were measured at the locations marked with a red star (P1), triangle (P2) and circle (P3) in (c). Abbreviations: hBN, hexagonal boron nitride; MATTG, magic-angle twisted trilayer graphene; TBG, twisted bilayer graphene; MG, monolayer graphene.

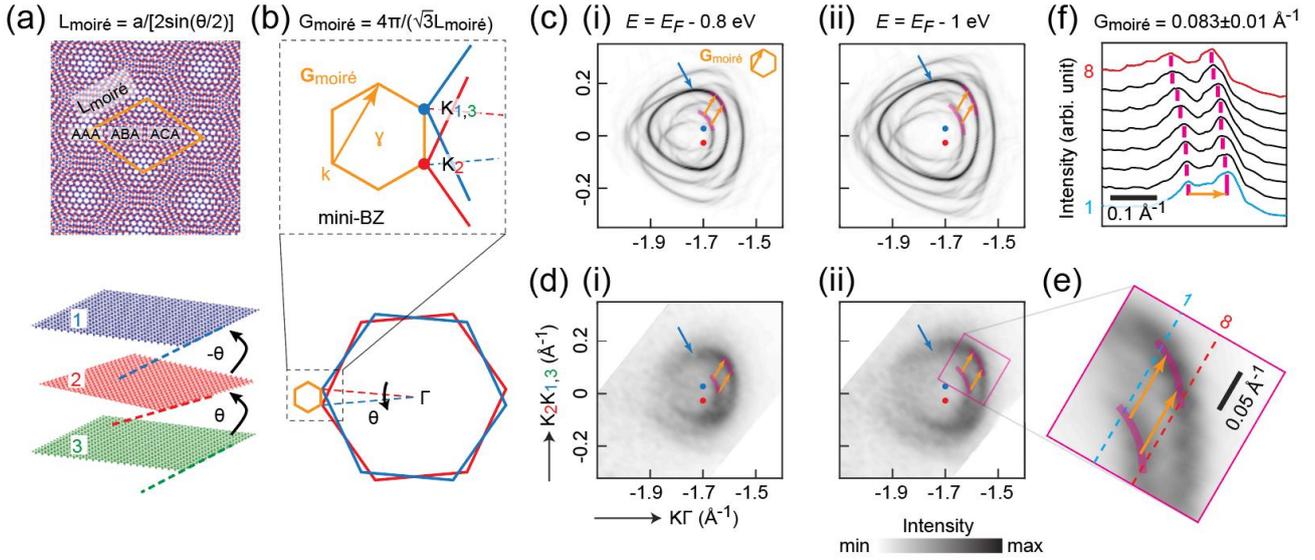

**Figure 2.** Band folding due to moiré lattice potential. a) Schematics of the stacking geometry of MATTG (lower panel) and the moiré pattern as seen from the top view (upper panel). The orange rhombus indicates moiré unit cell. b) The mini-Brillouin zone (mini-BZ) is constructed from the difference between $K_{1,3}$ and $K_2$ of the three layers (lower panel). The zoom-in view of the mini-BZ is shown in the upper panel. The blue, red and orange hexagons indicate the BZ of the top-layer (bottom-layer) graphene, middle-layer graphene and mini-BZ, respectively. c,d) Calculations of the spectral weight within the continuum model (c), and corresponding ARPES constant energy contours (d) at -0.8 eV and -1 eV, respectively. e) A zoom-in view of the area indicated by the magenta square in Figure 2d(ii). f) The MDCs between the momenta indicated in (e). $K_{1,3}$ and $K_2$ are marked by blue and red dots in (b)-(d). The orange arrows in (b) to (e) indicate the reciprocal vectors of the mini-BZ. The blue arrows indicate the gapless Dirac cone in (c) and (d). The magenta lines are guides to the eye of spectral weight extrema in Figure 2c-f

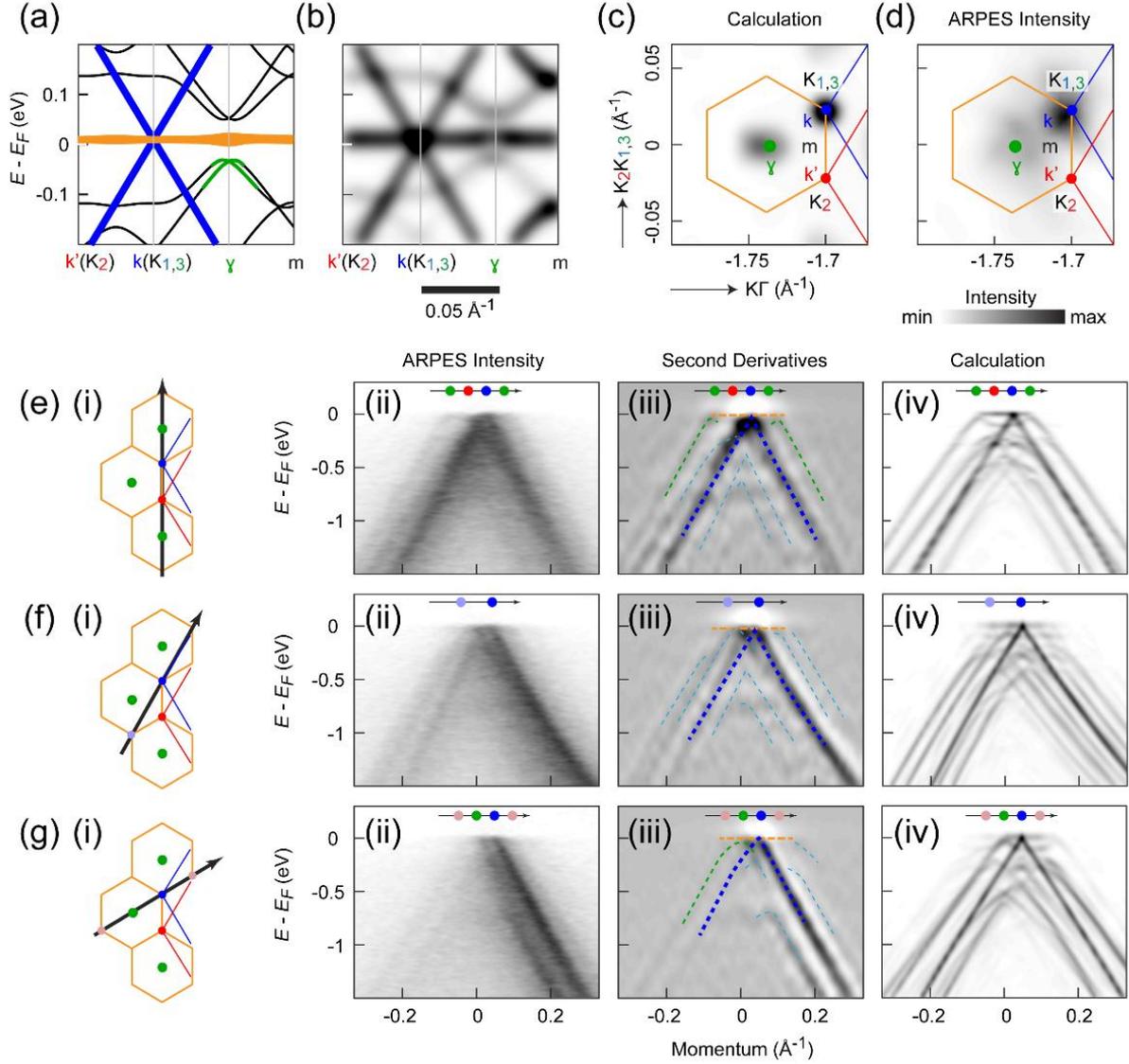

**Figure 3**. Visualization of the moiré mini-bands, flat bands and Dirac bands. a,b) Calculated band structure in the mini-BZ (a) and calculated spectral weight within the continuum model near $K_{1,3}$ and $K_2$ (b). The blue, orange and green lines in (a) indicate the gapless Dirac band, flat bands and the band top at γ, respectively. c,d) Calculation of the spectral weight (c) and corresponding ARPES contours at the Fermi energy (d) with an energy window of 50 meV. e-g) Dispersion plot of ARPES (ii), second derivatives (iii) and spectral weight calculation (iv) along the momentum directions indicated in (i). In (c) to (g), the orange hexagons mark the mini-BZs, and the green, blue and red dots mark the high symmetry point γ, k ($K_{1,3}$) and k' ($K_2$), respectively. In Figure 3e-g(iii), the blue, orange, green and cyan dashed lines are guides of the eye of the Dirac bands, flat bands, the band top at γ and moiré mini-bands, respectively. The spectral weight calculation is convoluted with Gaussian function along momentum and energy axis ($\Delta k = 0.01 Å^{-1}, \Delta E = 35\ meV$).

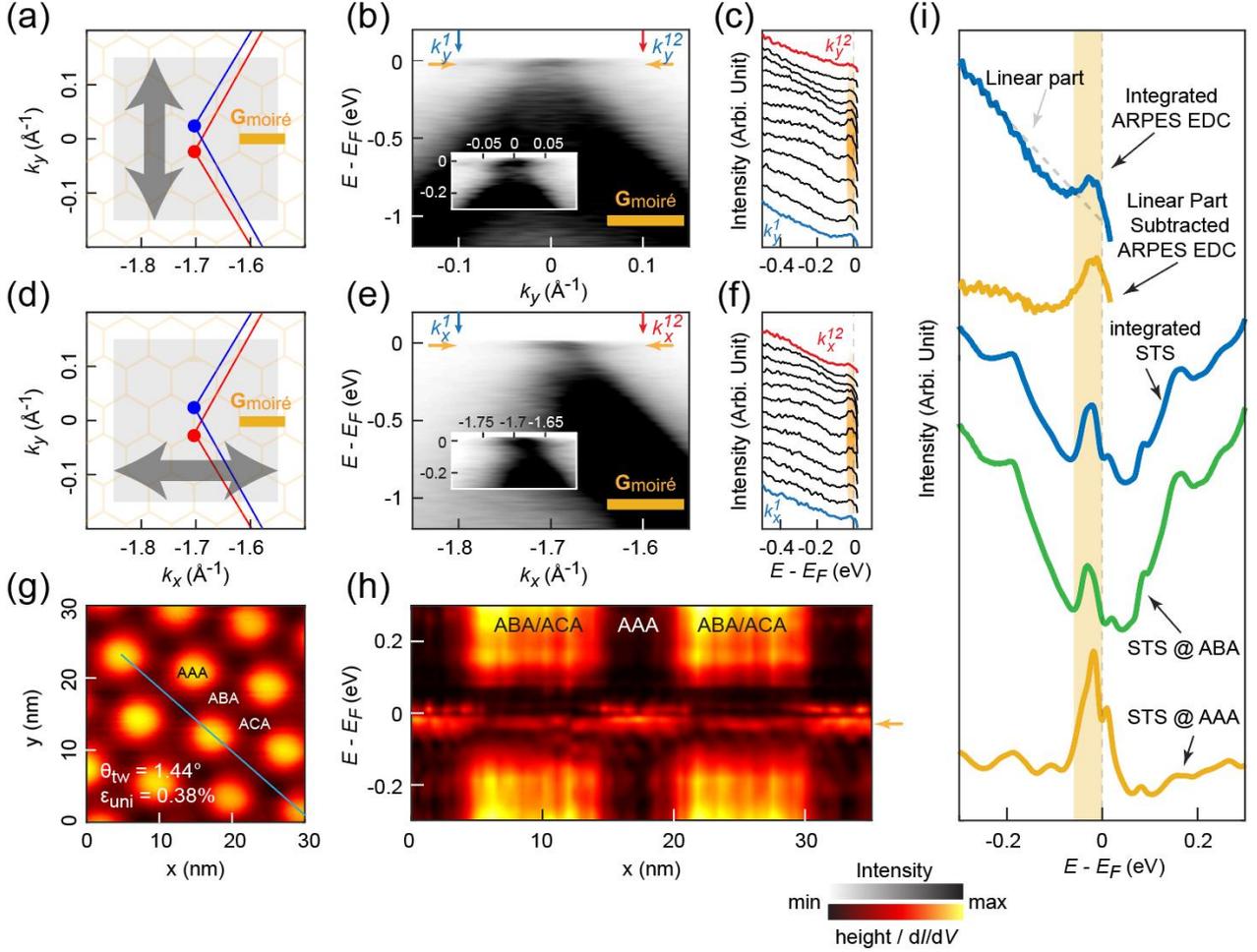

**Figure 4**. Moiré flat bands visualization in momentum and real space. a,b) Integrated ARPES dispersion plot (b) along $k_y$ direction in the area as indicated by the grey shaded square in (a). Inset: zoom-in plot showing the flat bands. c) EDCs between the momenta indicated in (b). d-f), Same as (a)-(c) but the dispersion plot is along $k_x$ direction. $k^1_{x(y)}$ and $k^{12}_{x(y)}$ are defined as $k^1_{x(y)} - k^D_{x(y)} = -0.1$Å and $k^{12}_{x(y)} - k^D_{x(y)} = 0.1$Å, where $k^D_x = -1.7$Å and $k^D_y = 0$ are the momentum coordinates of the Dirac point. $k^1_{x(y)}$ to $k^{12}_{x(y)}$ are evenly spaced. In (a) and (d), orange hexagons mark the mini-BZs. in (c) and (f), orange shading areas are flat bands. g,h) STS line cut (h) along the blue line marked in the STM topography (g) across alternating ABA, ACA and AAA regions. In (g), the twist angle and strain are indicated (see Figure S6, Supporting Information). In (b), (e) and (h), the orange arrows indicate the lower flat band. Set voltage $V_{set}$ = -140 mV, set current $I_{set}$ = 100 pA and modulation voltage $V_{mod}$ = 5 mV. i) Comparisons among the integrated ARPES EDC in the momentum area marked by the grey shaded square in (a) and (d), linear part subtracted ARPES EDC, integrated STS and STS collected at ABA and AAA regions. The shaded area indicates the energy window of the lower flat band.